\documentclass[iop,12pt,a4paper,showpacs,epsfigure,superscriptaddress]{iopart}

\usepackage{hyperref}
\usepackage{graphics}
\usepackage{amsfonts}
\usepackage{bbm}
\usepackage{color,epstopdf}
\usepackage{ulem,bm}
\usepackage{cancel}
\usepackage{amsbsy,iopams}

\newcommand{\ket}[1]{\vert #1 \rangle}

\newcommand{\ketbra}[2]{\vert #1 \rangle \langle #2 \vert}

\begin{document}

\title{Squeezing of mechanical motion via qubit-assisted control}

\author{Marco G. Genoni}
\address{Department of Physics \& Astronomy, University College London, 
Gower Street, London WC1E 6BT, United Kingdom}
\address{QOLS, Blackett Laboratory, Imperial College London, London SW7 2BW, UK}

\author{Matteo Bina}
\address{Dipartimento di Fisica, Universita' degli Studi di Milano, I-20133 Milano, Italy}

\author{Stefano Olivares}
\address{Dipartimento di Fisica, Universit\`a degli Studi di Milano, I-20133 Milano, Italy}
\address{CNISM UdR Milano Statale, I-20133 Milano, Italy}

\author{Gabriele De Chiara}
\address{Centre for Theoretical Atomic, Molecular and Optical Physics, School of Mathematics and Physics, Queen's University, Belfast BT7 1NN, United Kingdom}

\author{Mauro Paternostro}
\address{Centre for Theoretical Atomic, Molecular and Optical Physics, School of Mathematics and Physics, Queen's University, Belfast BT7 1NN, United Kingdom}


\date{\today}

\begin{abstract} 
We propose a feedback control mechanism for the squeezing of the phononic mode of a mechanical oscillator. We show how, under appropriate working conditions, a simple adiabatic approach is able to induce mechanical squeezing. We then go beyond the limitations of such a working point and demonstrate the stationary squeezing induced by using repeated measurements and re-initialisation of the state of a two-level system ancilla coupled to the oscillator. Our non-adaptive feedback loop offers interesting possibilities for quantum state engineering and steering in open-system scenarios.
\end{abstract}


\maketitle


The development of the first generation of devices based on the paradigm of quantum technology requires the design of feasible schemes for quantum control. A considerable body of work has been recently produced in this sense~\cite{varie} and a few significant test-bed demonstration have been reported. Proposals for the fast cooling of the vibrations of trapped ions and micro-mechanical oscillators based on simple controlling schemes~\cite{Cerrillo} have been put forward recently. Moreover, techniques for the achievement of quantum optimal control have been extended to the dynamics of quantum many-body systems~\cite{OCT}.

 However, a number of hurdles are clearly on the route towards the full grounding of such schemes, ranging from strong environmental effects to the difficulty of addressing directly fragile quantum systems. Such challenges are even more important for devices exploiting mesoscopic systems, which display enhanced sensitivity to environmental decoherence. 

An architecture that seems to offer a chance to bypass such hindrances combines simple (effective) spin systems and vibrating micro or nano-structures~\cite{Treutlein} and aims at building hybrid devices of enhanced flexibility (thanks to the possibility of tuning the mutual coupling strengths amongst the various parts of the system) and robustness (enforced by the possibility to address the spin subsystem without affecting the oscillator)~\cite{reviewbelfast}. Interesting experimental demonstrations have been performed in this sense~\cite{Arcizet}, 
and recent endeavours have shown the possibility to engineer mechanisms able to enforce non-classical features in massive mechanical systems~\cite{Vanner}. Yet, the route towards the consolidation of such methods is still long.  

Here we contribute to the aforementioned quest by presenting a scheme that exploits a `hybrid' architecture of the form sketched above to achieve large squeezing of a harmonic oscillator via a feedback-assisted protocol built on repeated projections of an ancillary qubit and its reinitialisation. We demonstrate significant steady-state squeezing in a wide range of operating regimes of the system. In particular, our scheme does not require the time-gated switching on/off of the qubit-oscillator interaction, and thus relaxes significantly the degree of control required for the implementation of the protocol that we propose. Our scheme is, in this working principles, very close to the current design of hybrid configurations for the control of quantum harmonic oscillators embodied by massive mechanical structures~\cite{reviewbelfast,Arcizet} and can be applied to superconducting microstrip resonators coupled to superconducting qubits, a scenario that might be useful for the achievement of large squeezing of itinerant microwave radiation~\cite{Wallraff}.

\section{Effective interaction models} 

We consider the coupling between a qubit and an oscillator regulated by the Hamiltonian model 
\begin{equation}
\label{H1}
\hat{\cal H}_1=({\omega_a}/{2})\hat{\sigma}_z +\omega_m ( \hat{a}^\dag\hat{a} +{1}/{2})+g\,\hat{\sigma}_x(\hat a+\hat a^\dag),
\end{equation} 
where we have assumed units such that $\hbar=1$ throughout the manuscript$, \omega_m$ is the frequency of the oscillator (with annihilation and creation operators $\hat a$ and $\hat a^\dag$), $\omega_a$ is the transition frequency between the levels $\{\ket{g},\ket{e}\}$ of the qubit, $g$ is the interaction strength, and $\hat\sigma_{j}$ is the $j=x,y,z$ Pauli matrix. Finally, we have introduced the slowly varying quadrature operator $\hat x_1\equiv\hat a+\hat a^\dag$, whose squeezing properties will be addressed here. This model can be physically embodied by a few systems, including the case of a mechanical resonator (endowed with a magnetic tip) coupled to a nitrogen-vacancy centre in diamond exposed to a strong transverse magnetic field~\cite{Rabl} or the interaction between a nano-mechanical resonator and a Cooper-pair box~\cite{Rabl2}. An alternative scenario is provided by an intra-cavity atom that interacts with an externally driven cavity mode. The latter is, in turn, coupled through radiation-pressure to the vibrational mode of a mechanical cavity end-mirror~\cite{vacanti}, as it is typical of cavity-optomechanical settings~\cite{OptomechReview}. In this context, Eq.~(\ref{H1}) would be achieved by assuming the bad-cavity limit and eliminating adiabatically the field mode so to obtain a direct coupling between the atom and the mechanical mode. All these systems offer wide tunability of the relevant parameters as well as the possibility to prepare the state of the qubit and read it out accurately. A further configuration would involve a superconducting quantum interference device in the charge regime coupled with in a microstrip resonator~\cite{Blais}. However, here we focus on mechanical bosonic systems for which the non-classical features we are interested in remain yet to be demonstrated experimentally. 


{We move to a rotating frame defined by the free qubit Hamiltonian $\hat{\cal H}_{\sf qubit} = \omega_a\hat{\sigma}_z/2$, obtaining   
\begin{eqnarray}
\hat{\cal H}_{1,{\sf int}}(t) &= \omega_m \: \hat{a}^\dag\hat{a}  +g e^{i\omega_a} \hat{\sigma}_+ \: \hat{x}_1  + g e^{-i\omega_a} \hat{\sigma}_-\: \hat{x}_1 \:.
\end{eqnarray}
As we consider the large detuning regime $\delta\equiv \omega_a-\omega_m \gg g$, we can 
average over the fast rotating terms and thus performing the adiabatic elimination of the qubit excitations as described in Ref.~\cite{GamelJames}. This procedure yields the effective Hamiltonian
\begin{eqnarray}\label{H2}
\hat{\cal H}_{\sf eff} &=& \hat{\cal H}_0 + \frac{1}{\omega_a} [ \hat{h}^\dagger,\hat{h}] = \omega_m \:  \hat{a}^\dag\hat{a} +\frac{g^2}{\omega_a}\,\hat{\sigma}_z \otimes \hat x^2_1 \:,
\end{eqnarray}
where we have defined $\hat{\cal H}_0 = \omega_m \: \hat{a}^\dagger \hat{a}$ and $\hat{h}=g \hat{\sigma}_- \: \hat{x}_1$.}
An alternative approach to the achievement of the very same effective model is the use of the Schrieffer-Wolff transformation $\hat{\cal S}=e^{i\frac{g}{2\omega_a}\hat\sigma_y\otimes(\hat a+\hat a^\dag)}$~\cite{Bravyi}. When applied to $\hat{\cal H}_1$, such transformation projects the qubit-oscillator dynamics in the low-lying energy subspace. In fact, by using the operator-expansion formula truncated to the second order in $g/\omega_a$ we get
\begin{eqnarray}
\label{WS}
\hat{\cal S}\hat{\cal H}_1\hat{\cal S}^\dag&\simeq&\hat{\cal H}_1+i\frac{g}{2\omega_a}[\hat\sigma_y,\hat\sigma_z](\hat a+\hat a^\dag)+i\frac{g\omega_m}{2\omega_a}\hat\sigma_y(\hat a-\hat a^\dag)\nonumber\\
&+&\frac{g^2}{\omega_a}\hat\sigma_z(\hat a+\hat a^\dag)^2+{\cal O}(g^2/\omega^2_a).
\end{eqnarray}
By ignoring highly oscillating terms we obtain the effective model in Eq.~(\ref{H2}). Notice that under the assumption of strong coupling $g\lesssim\omega_a$ between qubit and harmonic oscillator, we shall retain the term containing $\hat x^2_1$~\cite{Niemczyk}. The presence, in such term, of $\hat a^2$ and $\hat a^{\dag2}$ suggests the possibility to enforce squeezing in the state of the oscillator. In what follows, we prove such intuition correct and carefully characterise the squeezing mechanism that we achieve. 

\section{Stabilizing the evolution} 

The mechanism embodied by Eq.~(\ref{H2}) would require a precise gating of the interaction between the qubit and the oscillator to achieve mechanical squeezing. Ideally, though, we would like to bypass such necessity and enforce non-classical features on the stationary state of the oscillator. To achieve this, we consider Hamiltonian $\hat{\cal H}_1$ and complement the interaction at hand with a dissipation channel, whose role is to stabilize the properties of the oscillator to steady-state conditions. 
{In order to keep our approach as general as possible, we consider the oscillator interacting with a phononic bath at finite temperature populated by $n_{\sf th}$ thermal phonons. The corresponding evolution is thus described by the master equation 
\begin{equation}
\label{ME}
\dot{\rho} = - i [\hat{\mathcal{H}_1},\rho] + \gamma (n_{\sf th} + 1) \mathcal{L}[\hat a]\rho +
\gamma n_{\sf th} \mathcal{L}[\hat a^\dagger]\rho
\end{equation}
with $\hat{\cal L}[\hat{A}]\rho=\hat{A} \rho\, \hat{A}^\dag-(\hat{A}^\dag \hat{A}\rho+\rho\hat{A}^\dag \hat{A})/2$ a trace-preserving Lindblad super-operator and $\gamma$ the coupling rate with the bath.}
To show that our approach is successful in achieving the anticipated squeezing, we consider the large-detuning limit, so that we can use the effective model $\hat{\cal H}_{\sf eff}$ instead of $\hat{\cal H}_1$ in Eq~(\ref{ME}) and carefully choose the initial preparation of the qubit. The intuition that we aim at exploiting consists in noticing that if the qubit is prepared in an eigenstate of $\hat\sigma_z$, we can replace $\hat{\sigma}_z\to\pm{1}$ in $\hat{\cal H}_{\sf eff}$ and thus achieve an effective Hamiltonian that affects only the harmonic oscillator and is quadratic in the relevant operators, thus ensuring the solvability of the dynamical equation. In line with such an intuitive approach, in the remainder of this work we consider the case of a qubit initially prepared in $\ket{e}$. 

Let us now address the solution of the dynamical model explicitly. The quadratic nature of the effective model discussed above and the assumption of an initial Gaussian state of the harmonic oscillator allow us to make use of the powerful framework of  Gaussian states. These are completely specified by their vector of first moments $\langle \hat{\bf r}\rangle$ and  covariance matrix (CM) ${\bm \sigma}$ whose elements are $\sigma_{jk}={\rm Tr}[ \{\hat{r}_j,\hat{r}_k\}\varrho\,]-2\,{\rm Tr} [\hat{r}_j \varrho\,]\, {\rm Tr}[\hat{r}_k \varrho\,]$, where $\varrho$ is the density matrix of the oscillator and $\hat{\bf r}=(\hat{x}_1,\hat{x}_2)^{\sf T}$ [with $\hat x_2=i(\hat a^\dag-\hat a$)] is the vector of the oscillator quadrature operators. The master equation can be converted into the following set of dynamical equations 
\begin{eqnarray}
\partial_t\langle \hat{\bf r} \rangle &=A\langle \hat{{\bf{r}}} \rangle, \nonumber \\
\partial_t{\bm \sigma} &= A\boldsymbol{\sigma}+{\bm \sigma} A^{\sf T}+ D, \label{eq:H2gauss}
\end{eqnarray}
where we have introduced the drift matrix $A=i\sigma_y {H}_{\sf eff} -\gamma {I}/2$ 
with $H_{\sf eff}$ the Hamiltonian matrix given by $\hat{\cal H}_{\sf eff}=\hat{{\bf{r}}}^{\sf T} H_{\sf eff} \,\hat{{\bf{r}}}/2$. The matrix $D=\gamma(2 n_{\sf th}+1) I$ with $I$ the identity matrix is responsible for diffusion. Equations similar to the one for ${\bm\sigma}$ which is of the well-known differential Lyapunov matrix form, are key for the study of the conditions for stability in control theory~\cite{Fang} and help addressing the dynamics of quantum systems subjected to open-loop and feedback-control mechanisms~\cite{GenoniSerafiniRogers}. 

It is physically reasonable and experimentally motivated to assume that the oscillator is initially at thermal equilibrium with its environment. This is the case, for instance, for micro- and nano-mechanical oscillators, which are typically fabricated on substrates sustaining spurious background phononic modes at a given temperature~\cite{Cole}. Needless to say, other experimentally motivated examples can be identified. We thus consider the initial thermal state $$\varrho(0)=\sum_n\frac{(\bar{m})^n}{(1+\bar{m})^{n+1}}\ketbra{n}{n}$$ with $\bar{m}=(e^{\beta\omega_m}-1)^{-1}$ average phonons of the oscillator, $\beta$ the inverse temperature, and $\ket{n}$ an element of the Fock basis. 
{Under such assumptions, we can analytically solve the differential equation for ${\bm \sigma}$, looking in particular for the steady state solutions. In the following we set $\bar{m}=n_{th}$ as the oscillator is in the equilibrium with his own bath described by Eq. (5). One can check that in the presence of dissipation (i.e. for $\gamma\neq 0$), the dynamical system is always stable as the sufficient condition $\lim_{t \to \infty} (e^{At}) =0$ is always satisfied. In this case, the oscillator reaches a steady state characterized by the following values of the
variances and covariance of the quadrature operators
\begin{eqnarray}
\Delta \hat{x}^2_1 = (1+2n_{\sf th})\left[ 1 - \frac{8g^2 \omega_m}{16 g^2 \omega_m + \omega_a (\gamma^2+4\omega_m^2)} \right], \label{eq:Deltax} \\
\Delta \hat{x}^2_2 = (1+2n_{\sf th})\left[1+\frac{32 g^4+8 g^2 \omega_a\omega_m }
{16g^2\omega_m\omega_a + \omega_a^2(\gamma^2+4\omega_m^2)}\right], \\
\Delta (\hat{x}_1\hat{x}_2) = -\frac{4g^2 \gamma(1+2n_{\sf th})}{16 g^2 \omega_m + \omega_a (\gamma^2+4\omega_m^2)}.
\end{eqnarray}
In the above equations $(1+2n_{\sf th})$ is the variance of the quadratures of a harmonic oscillator prepared in a thermal state and detached from the ancilla (that is, for  $g=0$). An example of the behavior of $\Delta \hat{x}^2_1$ and $\Delta \hat{x}^2_2$ against time and for $n_{\sf th}=0$ is reported in Fig.~\ref{f:DeltaxDeltap}. 
As it can be seen by inspecting the first of Eqs.~(\ref{eq:Deltax}), for $n_{\sf th}=0$ quantum squeezing of the $\hat{x}_1$ quadrature (i.e. $\Delta \hat{x}^2_1<1$) is achieved for any $g>0$. At non-zero temperatures, $\Delta \hat{x}^2_1$ is reduced with respect to the variance of a thermal state, thus showing noise reduction below the corresponding thermal shot noise. More explicitly
\begin{equation}
\Delta \hat{x}^2_R = \frac{\Delta \hat{x}^2_1}{1+2 n_{\sf th}} = 1 - \frac{8g^2 \omega_m}{16 g^2 \omega_m + \omega_a (\gamma^2+4\omega_m^2)}<1.
\end{equation}
In analogy with what is found for mechanical systems at the quantum level (cf. Refs.~\cite{ThermoSqBowen,Nunnenkamp}) we will refer to such effect as {thermomechanical squeezing}. Quite remarkably, such effect does not depend on the actual value of $g$ and is achieved for any non-null value of such parameter, thus showing the inherent efficiency of the protocol proposed herein.
\begin{figure}[t!]
\centering
\resizebox{.50\columnwidth}{!}{\includegraphics{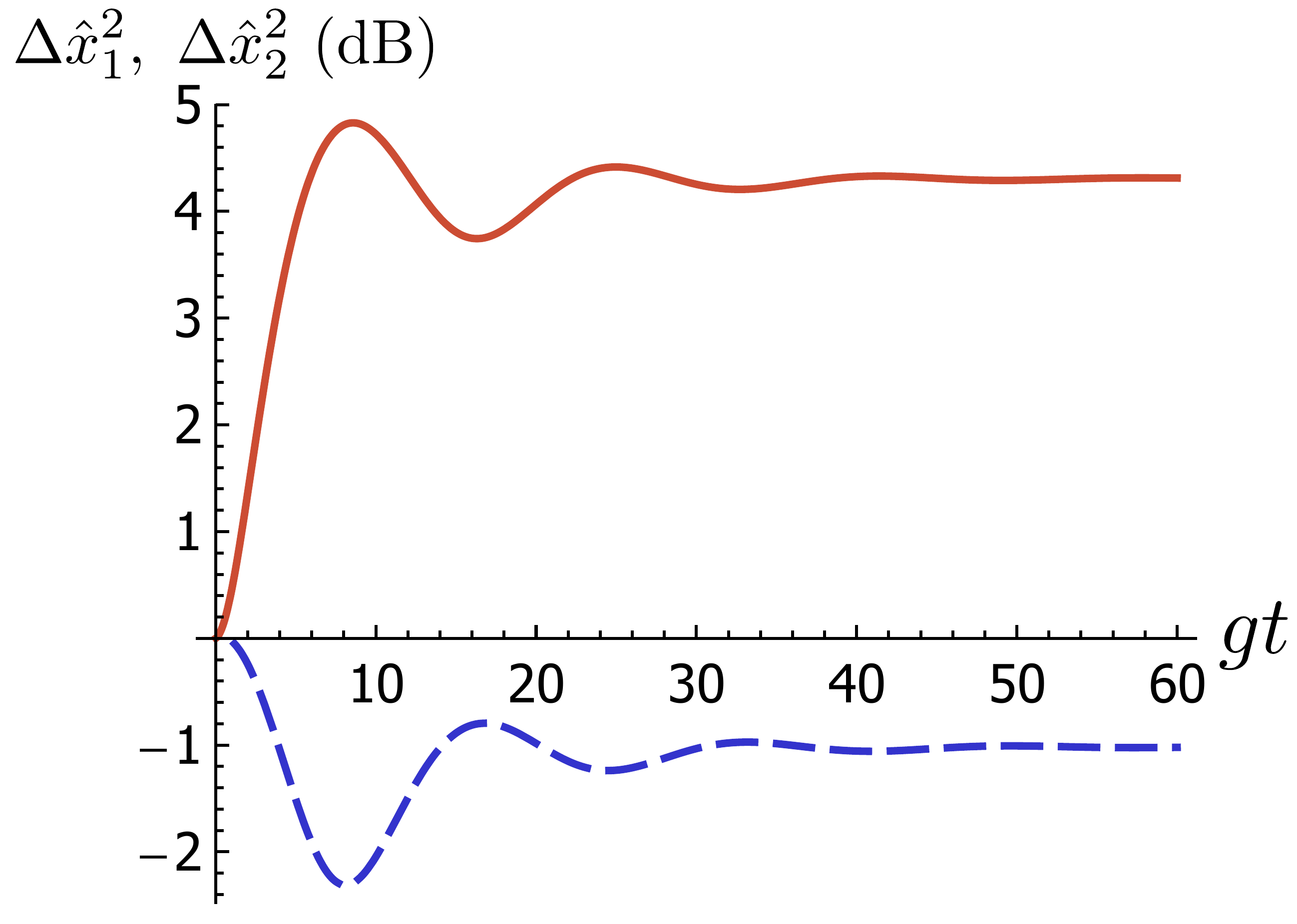}}
\caption{\label{f:DeltaxDeltap} Time evolution of the two variances $\Delta\hat{x}^2_1$ (blue dashed line) and $\Delta\hat{x}^2_2$ (red solid line) in dB-scale for the harmonic oscillator initialized in the vacuum state with $\omega_m=0.1 g$, $\gamma=0.1 g$ and $\omega_a=15g$.}
\end{figure}
}

\section{Numerical simulation of the ideal model} 

As it was made clear throughout its derivation, Eq.~(\ref{eq:Deltax}) depends crucially on the validity of the performed adiabatic elimination and the ability to keep the qubit in the state it has been initially prepared into $\ket{e}$ throughout the evolution. Such a possibility is not certain as far as model $\hat{\cal H}_1$ is concerned, although we expect that for large values of $\delta$ such a condition is met with good accuracy. The scope of our analysis herein is to test such expectations in a measurable way. 
\begin{figure}[htb!]
\centering
\centering{\bf (a)}\hskip6cm{\bf (b)}\\
\resizebox{.35\columnwidth}{!}{\includegraphics{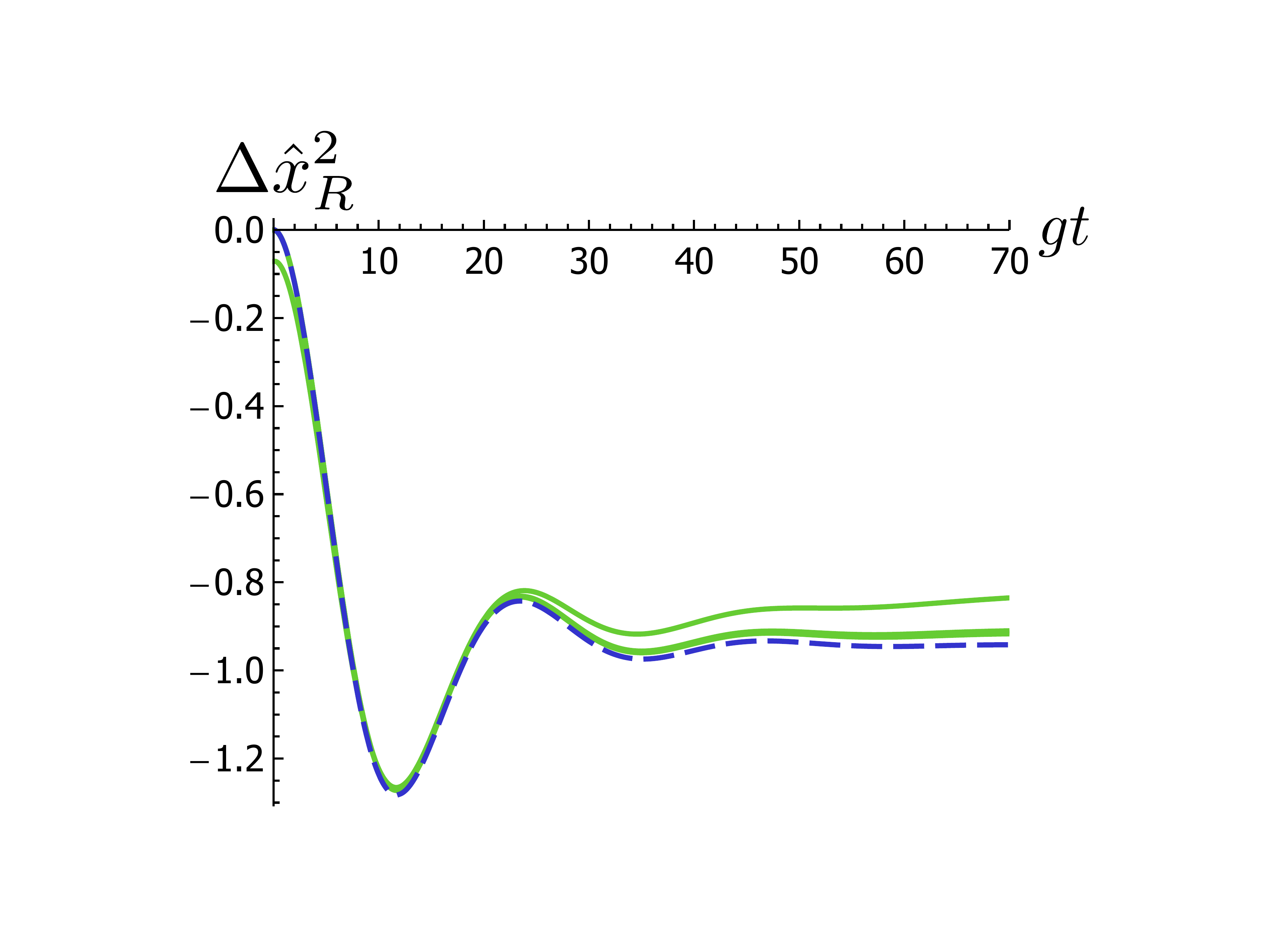} }
\resizebox{.355\columnwidth}{!}{\includegraphics{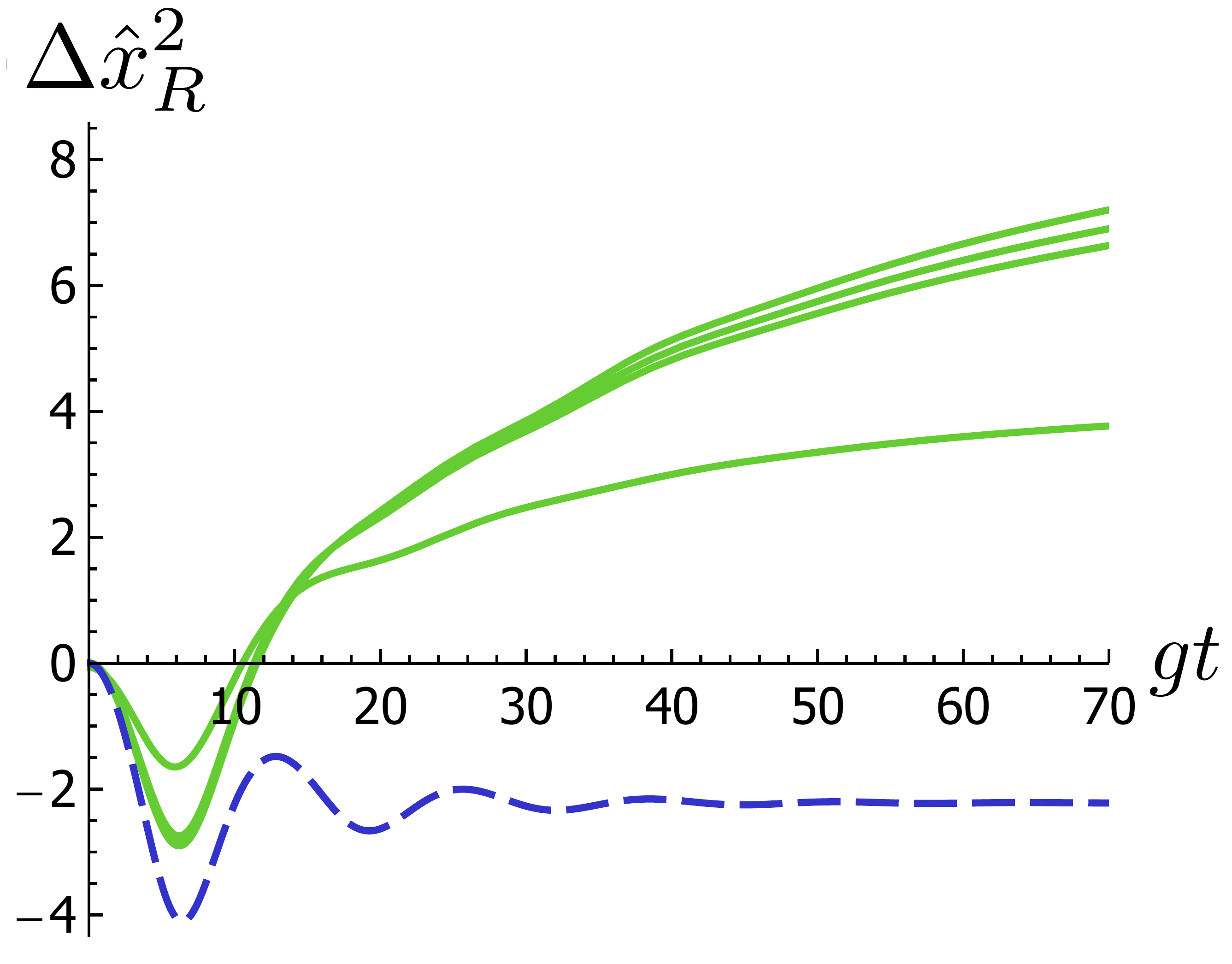} }
\caption{\label{f:detunings} Time evolution of the renormalized variance $\Delta \hat{x}_R^2$ in dB-scale. The harmonic oscillator is initialized in a thermal state having $\bar{m}=n_{\sf th}$ thermal phonons, we consider $\omega_m=\gamma=0.1 g$ and two different choices of $\omega_a$: $\omega_a = 50g$ [panel {\bf (a)}], and $\omega_a =8g$ [panel {\bf (b)}]. The solid green curves correspond to the numerical simulations with different average number of thermal phonons (from bottom to top: $n_{\sf th}=0.2, 0.3, 0.4, 3.0$), while the dashed blue curves correspond to the effective evolution governed by $\hat {\cal H}_{\sf eff}$ (which is insensitive to $n_{\sf th}$).}
\end{figure}

We thus proceed to fully simulate the evolution guided by $\hat{\cal H}_1$ and compare the corresponding results for the squeezing of the harmonic oscillator to the analogous quantity achieved using Eq.~(\ref{eq:Deltax}). As shown in Fig.~\ref{f:detunings}, it is indeed the case that a large value of $\delta$ results in values of $\Delta \hat{x}^2_R$ very close to the degree of squeezing achieved via the true dynamics. Quantitatively, we find a degree of squeezing of about $1$ dB for $gt\gtrsim 50$ [cf. Fig.~\ref{f:detunings} {\bf (a)}]. {While the agreement between the two predictions is perfect as far as $n_{\sf th}=0$, the increasingly thermal nature of the initial state of the harmonic oscillator results in only very small differences in the long-time values of $\Delta\hat x_R$ (we remind that $\hat {\cal H}_{\sf eff}$ is insensitive to $n_{\sf th}$)}. 

Somehow expectedly, by relaxing the assumption of large detuning we significantly worsen the performance of the protocol and considerable deviations from the ideal results are found.  Indeed, moderate or small values of $\delta$ favour transitions between the two logical states of the qubit, thus making the basic assumption on top of which our effective scheme is built (the qubit should remain in state $\ket{e}$ throughout the whole evolution) no longer tenable. As a consequence, a $\delta$-dependent threshold value of $gt$ exists starting from which we do not observe any squeezing. Unfortunately, this holds also for the case reported in Fig.~\ref{f:detunings}. Therefore, in order to enforce squeezing in the steady-state of the oscillator we need to implement some additional form of control. The description of such mechanism is the focus of the next Section. 

\noindent

\section{Feedback-loop mechanism for steady-state squeezing}

In order to effectively force the qubit to remain in its initial state, we rely on the implementation of a feedback-loop scheme based on the repeated measurement of the qubit's energy and its conditional projection on $\ket{e}$. More specifically, our feedback-assisted scheme can be described as follows: 
\begin{itemize}
\item We call $\rho(t_0)$ the state of the qubit-oscillator system at a given time $t_0$, 
and ${\Phi}_{\Delta t}$ the dissipative map [with Hamiltonian part given by Eq.~(\ref{H2})] describing its evolution within an interval $\Delta t$. 
\item At time  $t_1=t_0 +\Delta t$, we measure the qubit in the $\{\ket{g},\ket{e}\}$ basis. 
\item If the outcome of the projection reveals a transition of the qubit to its logical ground state $\ket{g}$, the spin-flip operation $\hat{\sigma}_x$ is applied on it. Otherwise, the system is evolved in time for another interval $\Delta t$. 
\end{itemize}
The average state of the system that arise from the application of the scheme above reads 
\begin{equation}
\label{eq:Rt1}
{\rho}(t_1) = p_e {\varrho}_e(t_1)  \otimes |e\rangle\langle e| + 
p_g {\varrho}_g(t_1) \otimes \hat{\sigma}_x |g\rangle\langle g| \hat{\sigma}_x,
\end{equation}
where ${\varrho}_k(t_1)={ \langle k | \Phi_{\Delta t}{\rho}(t_0) | k\rangle} /{p_k} $ is the conditional state of the oscillator when  the qubit is found in state $\ket{k}~(k=g,e)$ and $p_k$ is the corresponding detection probability. The protocol described above is then iterative until the oscillator reaches a steady state at which the variance of the $\hat x_1$ quadrature stabilises around a dynamics-dependent value. A scheme close in spirit to ours has been implemented to prepare a microwave radiation field in a Fock state~\cite{Haroche}.

A few comments are in order. First, it should be clear that the choice of $\Delta t$ is important for the success of the scheme. Its value results from the delicate trade-off between the intuitive necessity to perform the qubit projective measurement as often as possible (so to maintain $p_e\simeq1$ and thus mimic faithfully the ideal behaviour that would arise from $\hat{\cal H}_{\sf eff}$) and the need to wait for enough time to let the squeezing build up. The latter request is due to the fact that the effective Hamiltonian $\hat{\cal H}_{\sf eff}$ results from a second-order process and is, thus, `slow' with respect to the natural timescales of the system. 

\begin{figure}[t]
\centering{\bf (a)}\hskip6cm{\bf (b)}\\
\resizebox{.35\columnwidth}{!}{\includegraphics{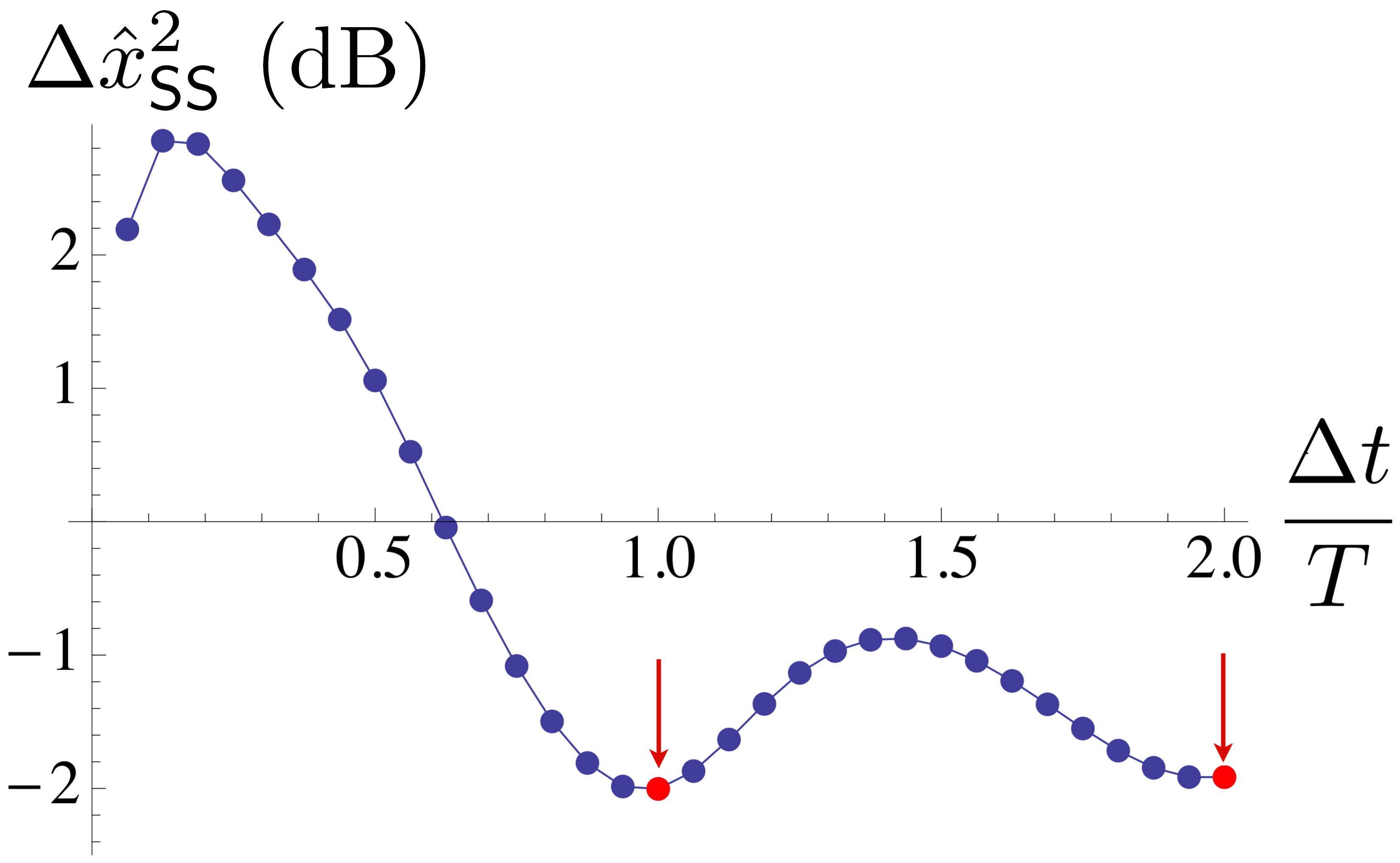} }
\resizebox{.35\columnwidth}{!}{\includegraphics{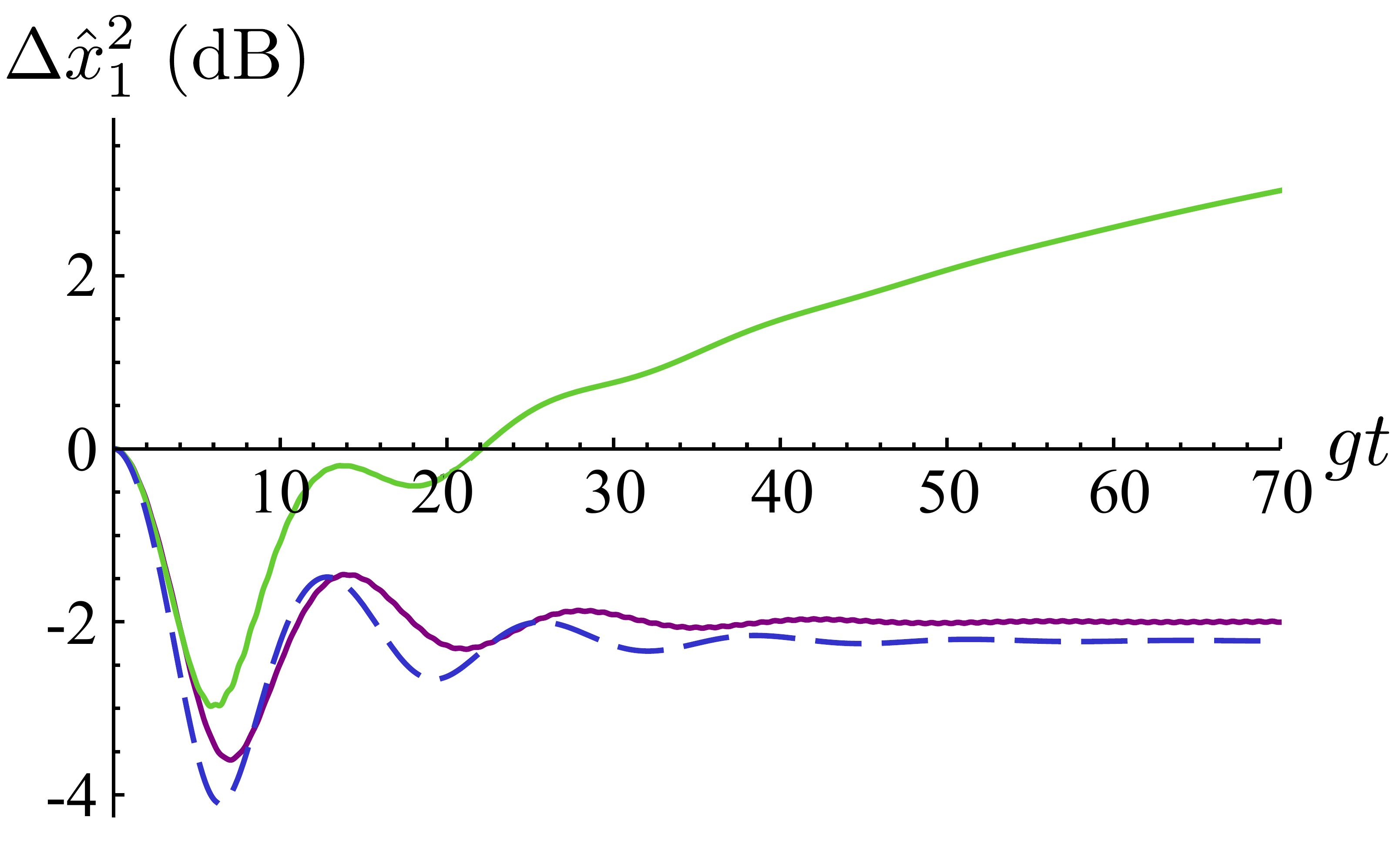} }
\caption{\label{Fig1} {\bf (a)}: Steady-state variance $\Delta \hat{x}^2_{\sf ss}$ obtained via the feedback protocol, against the corresponding time-step  $\Delta t$ written as a fraction of the two-level system period $T\equiv 2\pi/\omega_a$. The points corresponding to $\Delta t=T,2T$ are highlighted as they allow for optimal squeezing. 
{\bf (b)}: Variance $\Delta\hat{x}^2_1$ in dB-scale against $g t$. The numerical simulation of the optimized feedback protocol (purple curve) agrees with the analytical solution described by the effective Hamiltonian $\hat {\cal H}_{\sf eff}$ (blue dashed curve).
The green curve, showing no steady-state squeezing, illustrates the results of the numerical simulation without feedback.
In both plots we have used $n_{\sf th}=0$, $\omega_a=8g$, $\omega_m=\gamma=0.1g$.
}
\end{figure}

Let us now characterise the performance of the protocol by first addressing the case of a zero-temperature bath (i.e. $n_{\sf th}=0$). In Fig.~\ref{Fig1} {\bf (a)} we report the value of the variance $\Delta\hat{x}^2_{\sf ss}$ of quadrature $\hat x_1$ at steady-state, against the time interval $\Delta t$. Clearly, the degree of squeezing is a non-monotonic function of $\Delta t$ that results in an oscillating behaviour. The minima of such function correspond to $\Delta t=2p\pi/\omega_a~(p\in{\mathbb Z})$, i.e. multiples of the time taken by the qubit to make a transition between its states. The choice of $p=1$ allows for the achievement of the largest degree of squeezing as a compromise between the coherent protocol and the dissipative mechanism. In the rest of our study we will assume $\Delta t = 2\pi/\omega_a$, even for the cases of $n_{\sf th}\neq0$. 

Having determined the optimal size of the time interval for the evolution, we now establish a performance-benchmark by comparing the ideal results that would arise from the dynamical Eqs.~(\ref{eq:H2gauss}) to the results obtained through the numerical simulations based on $\hat{\cal H}_1$ and those arising from the implementation of the feedback-loop protocol optimised as discussed above. In Fig.~\ref{Fig1} {\bf (b)} we show that the feedback-assisted protocol reproduces closely the evolution induced by the effective model in Eq.~(\ref{H2}), resulting in a degree of squeezing at the steady state that is comparable to the value achieved via Eq.~(\ref{eq:Deltax}). As expected, no steady-state squeezing is achieved if no feedback is implemented. We thus conclude that the mechanism implemented throughout the feedback-assisted protocol is indeed able to closely resemble the desired effective squeezing Hamiltonian, at least for the case of a zero-temperature bath. 

\begin{figure}[t]
\centering
\hskip0.7cm{\bf (a)}\hskip5cm{\bf (b)}\hskip5cm{\bf (c)}\\
\resizebox{.3\columnwidth}{!}{\includegraphics{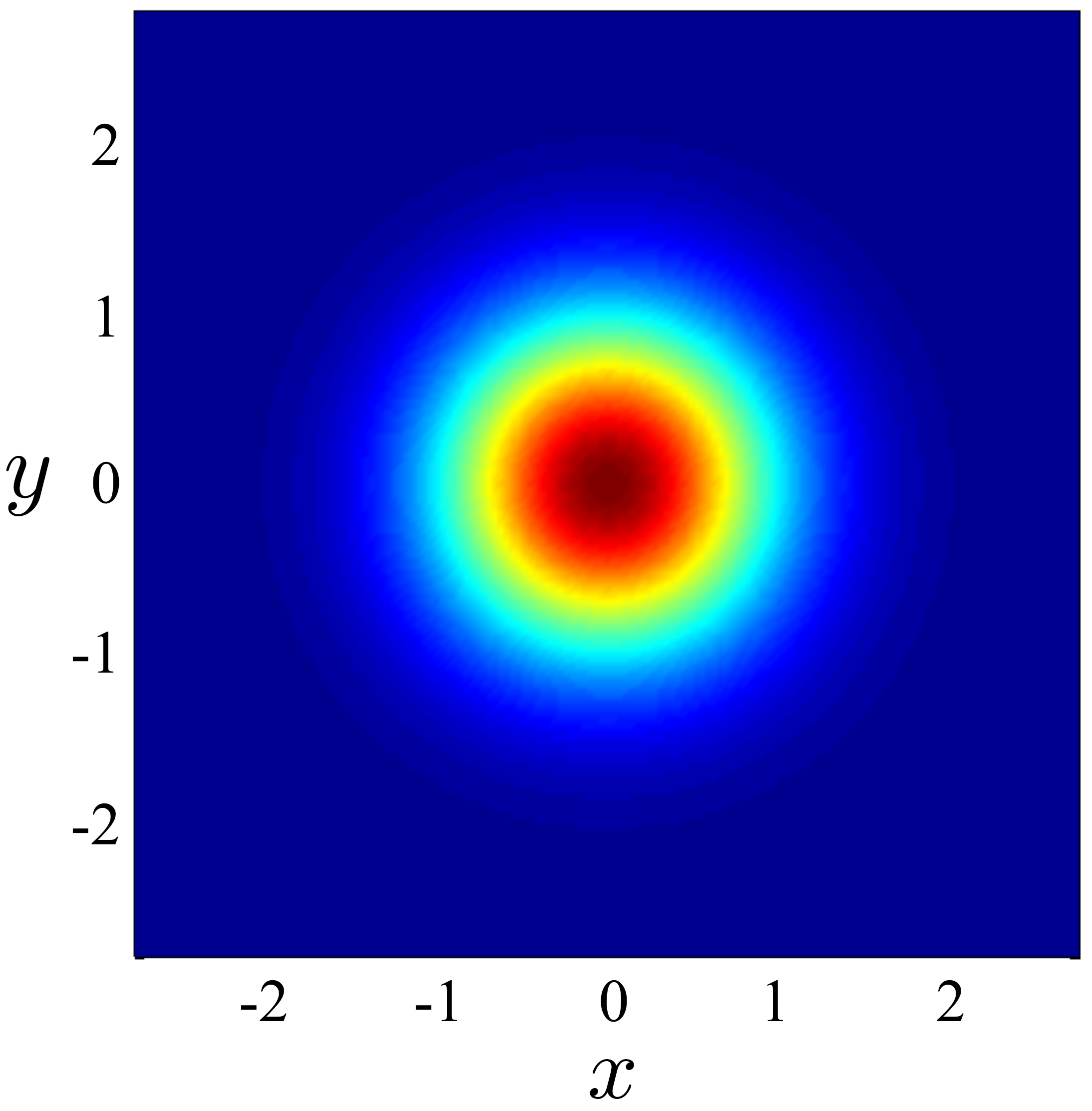} }
\resizebox{.3\columnwidth}{!}{\includegraphics{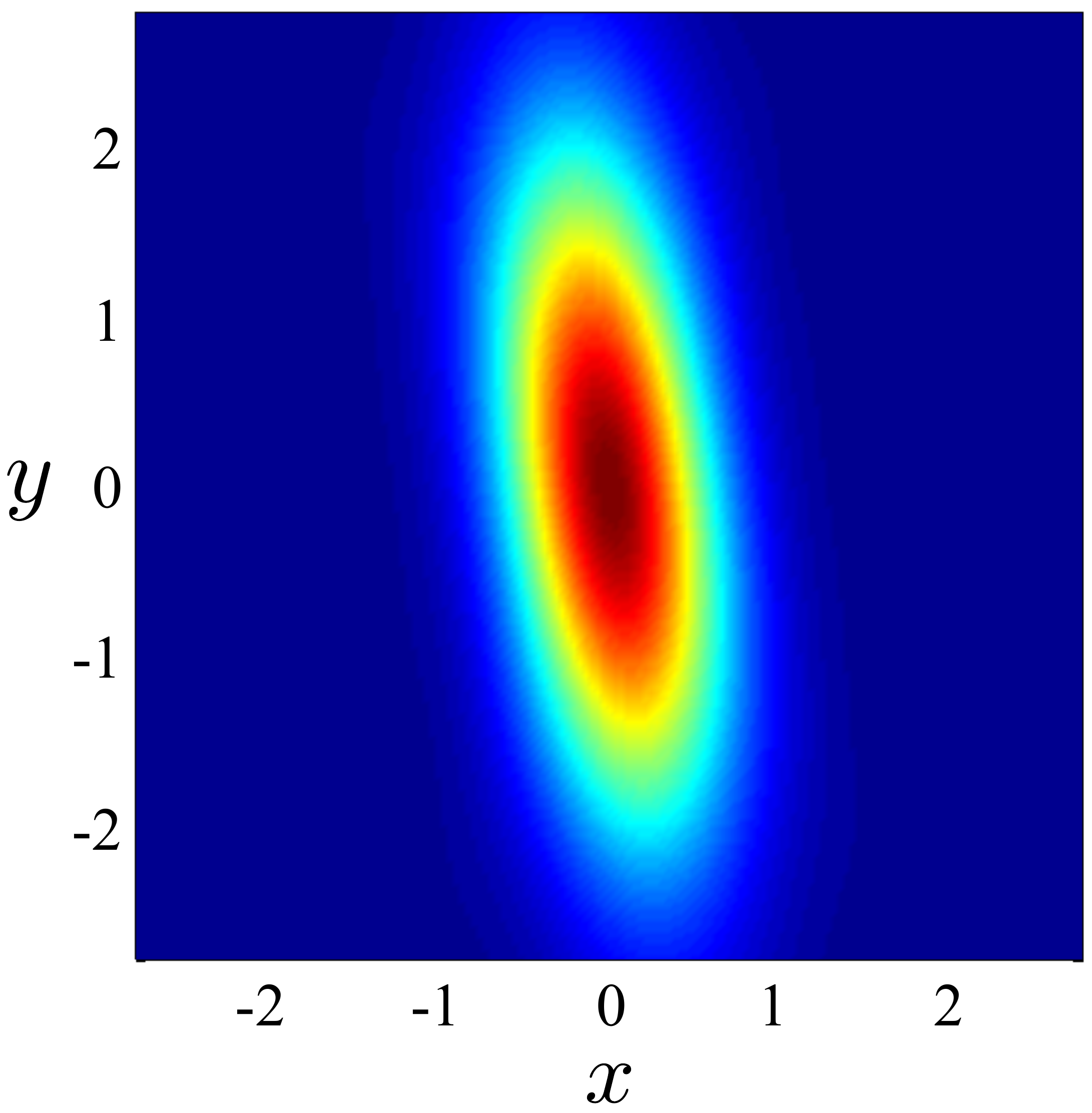} }
\resizebox{.3\columnwidth}{!}{\includegraphics{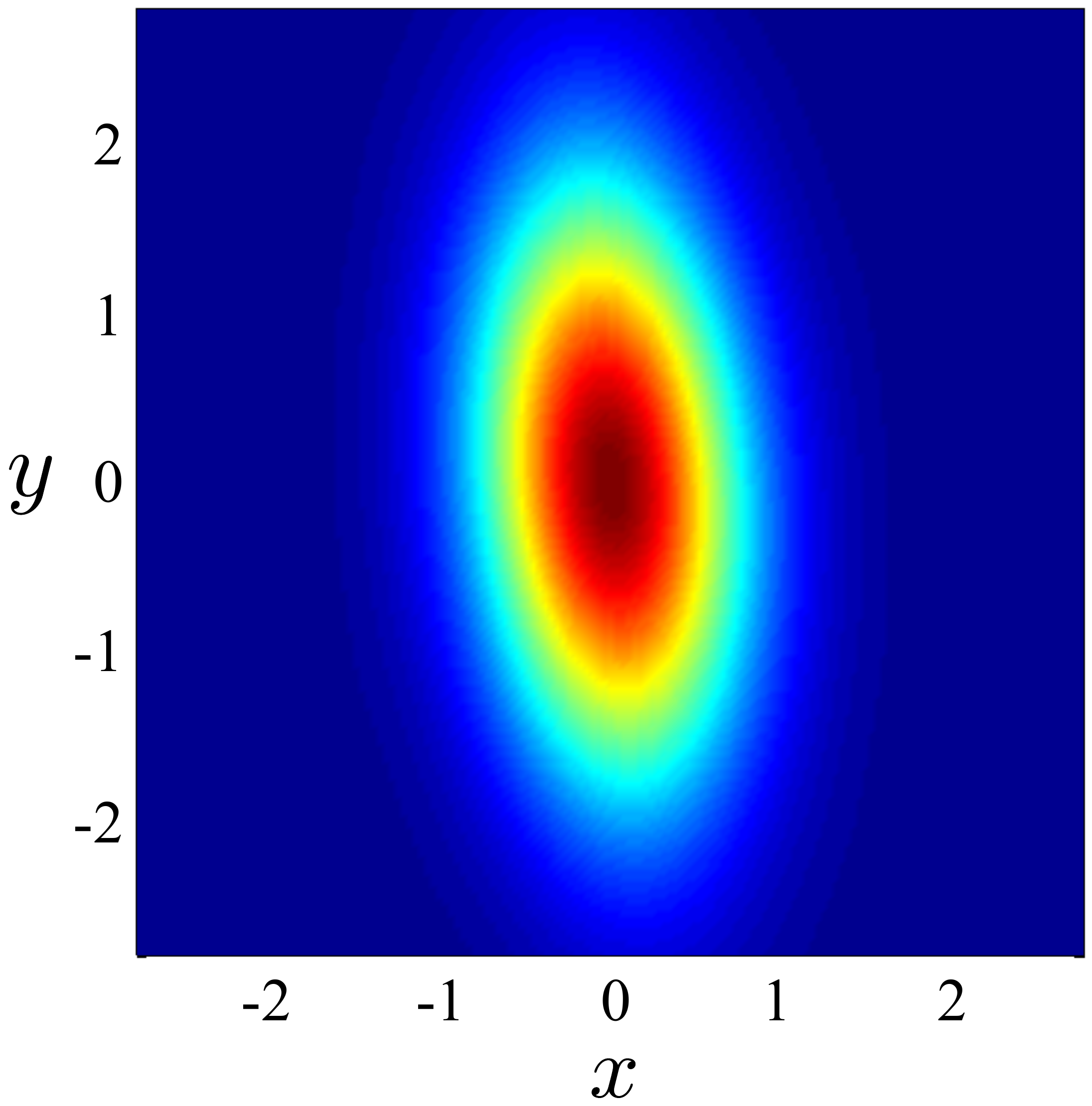}}
\caption{Snapshots of the evolution of the Wigner function corresponding to the state of the harmonic oscillator for $n_{\sf th}=0$, $\omega_a=8g$, $\omega_m=\gamma=0.1g$ and $gt=0$ [panel {\bf (a)}], $7$ [{\bf (b)}], and $70$ [{\bf (c)}].}
\label{wigner}
\end{figure}

\begin{figure}[t!]
\centering
\resizebox{0.6\columnwidth}{!}{\includegraphics{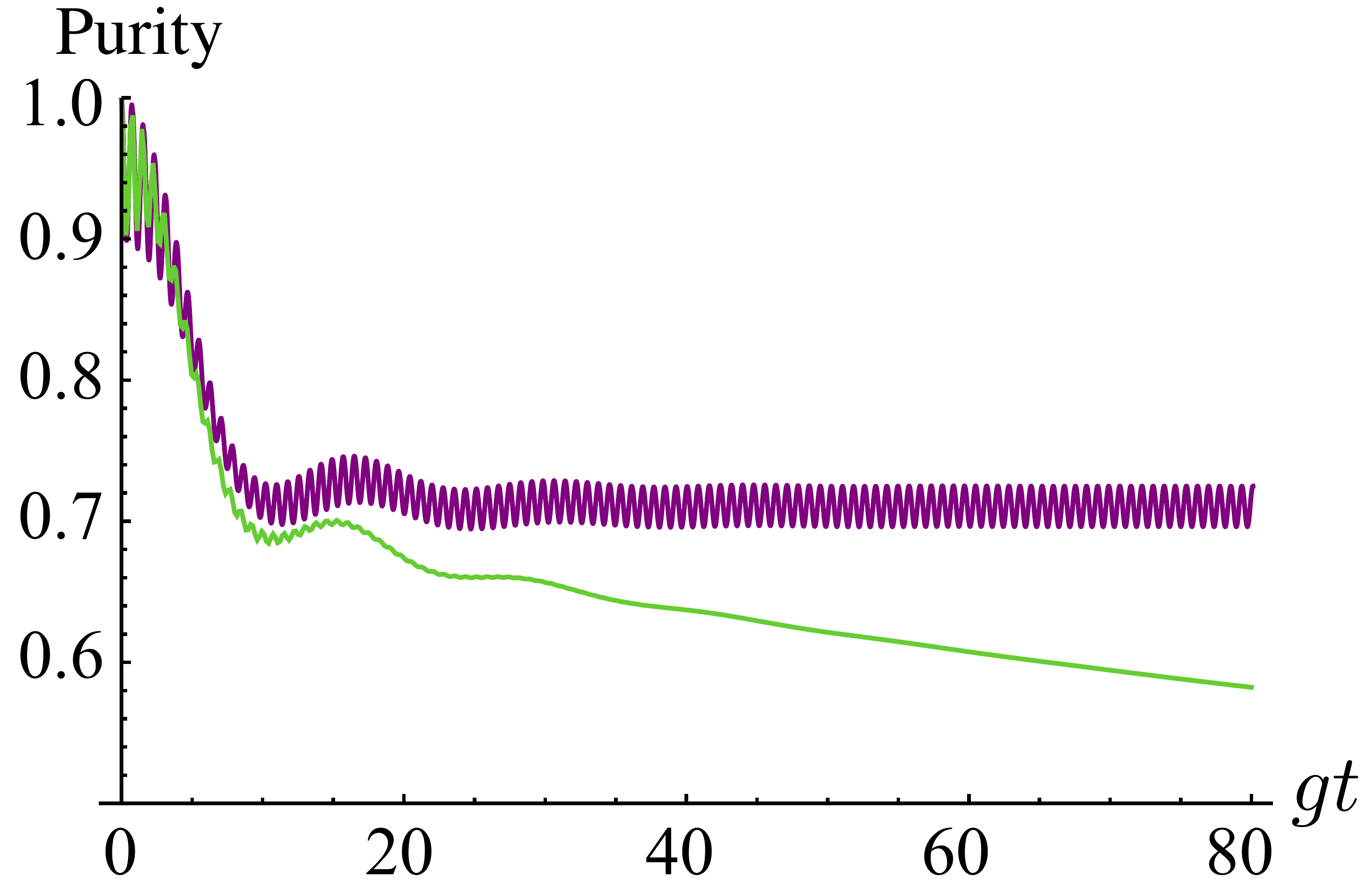} }
\caption{\label{PurityWig} We plot the purity of the state of the oscillator against the dimensionless interaction time $gt$. The green (purple) line is for the dynamics of the oscillator without (with) feedback-assited protocol. Other parameters are the same as in Fig.~\ref{wigner}.}
\end{figure}

Before moving to the assessment of the case with $n_{\sf th}\neq0$, we aim at providing further insight into the phenomenology of the squeezing process implemented through our qubit-assisted protocol. In order to do so, in Fig.~\ref{wigner} we show snapshots of the evolution of the Wigner function  
\begin{equation}
\label{wignerfunction}
W(x,y,t)=\frac{1}{\pi^2}\int {\rm Tr}[\rho(t)\hat D(\alpha)]e^{-2i(x\alpha_i-y\alpha_r)}d^2\alpha
\end{equation}
associated with the state of the harmonic oscillator. Here $\hat D(\alpha)=e^{\alpha\hat a^\dag-\alpha^*\hat a}$ is the displacement operator of amplitude $\alpha=\alpha_r+i\alpha_i$. As $gt$ grows, clearly squeezing builds up starting from the initial vacuum state, as seen from the evident anisotropy of the Wigner function. To illustrate such effect, we have picked up a few significant instants of time. Panel {\bf (b)} shows the Wigner function corresponding to the first minimum displayed in the purple curve in Fig.~\ref{Fig1} {\bf (b)}. Panel {\bf (c)} is for $gt=70$, when the dynamical degree of squeezing is the same as at the steady state. In Fig.~\ref{PurityWig} we compare the purity of the oscillator when the feedback protocol is implemented with what is achieved in the absence of it. Clearly, the steady state of the oscillator has a higher purity when its evolution is assisted by the re-initialised two-level system.  
Therefore, this analysis reinforces the idea that the feedback assisted protocol that we have devised progressively projects the state of the harmonic oscillator onto a high-purity squeezed state.  

\begin{figure}[b!]
\centering
\centering{\bf (a)}\hskip6cm{\bf (b)}\\
\resizebox{0.36\columnwidth}{!}{\includegraphics{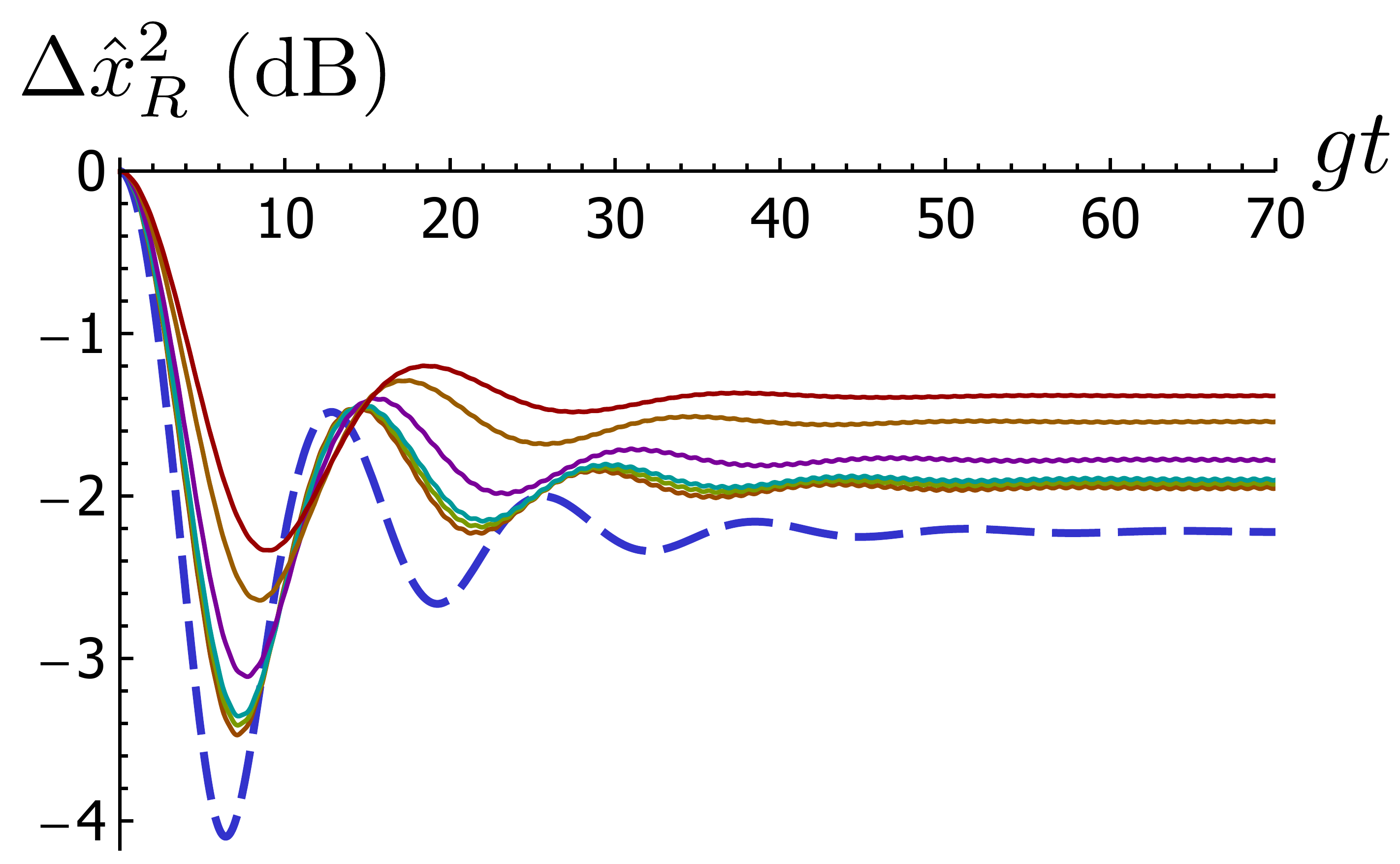} }
\resizebox{0.35\columnwidth}{!}{\includegraphics{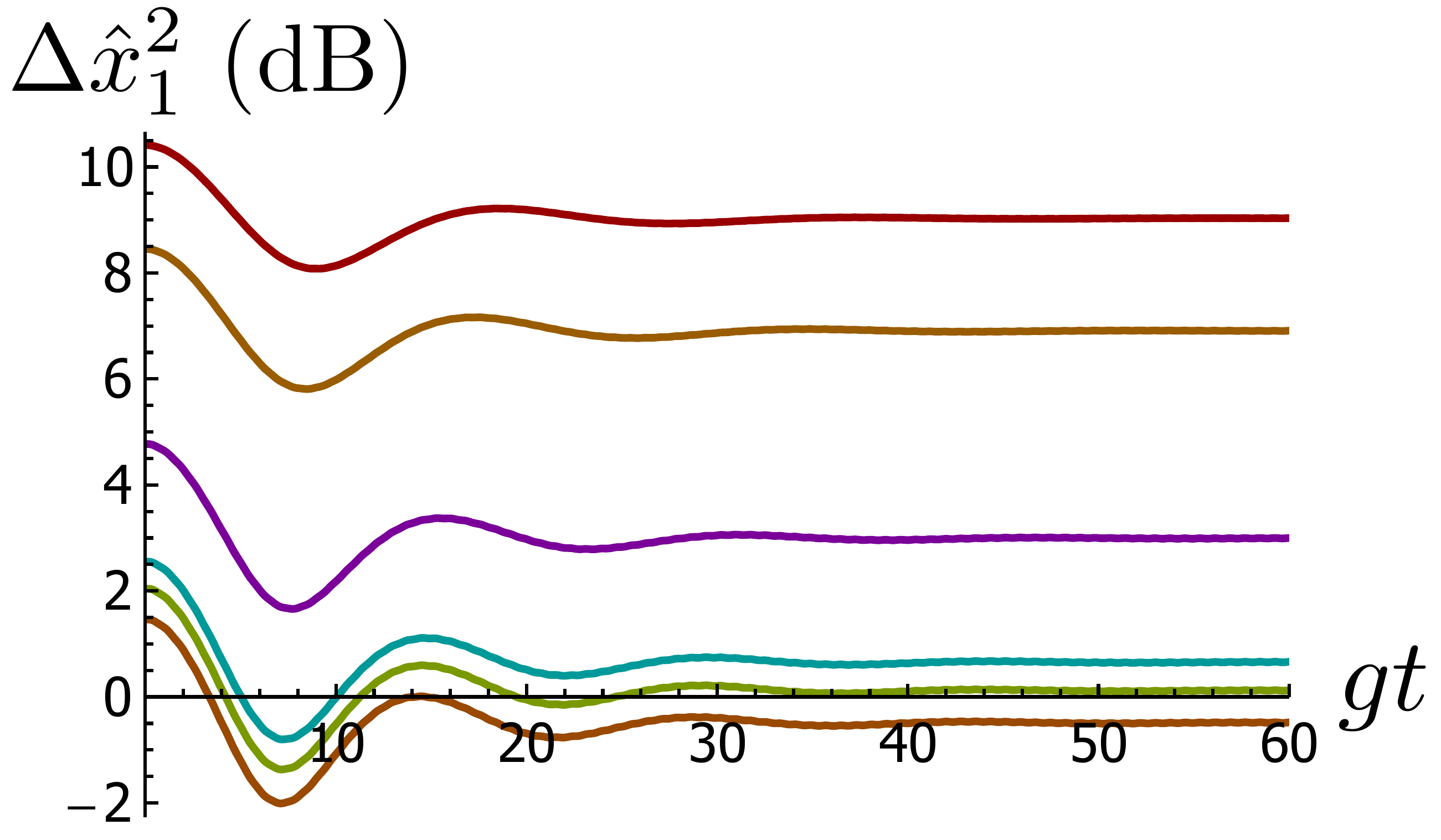} }
\caption{
 \label{ThermSqFeed} 
 {\bf (a)}: Time evolution of the renormalised variance $\Delta\hat{x}_R^2$ in dB-scale when the feedback protocol is implemented and 
with $\omega_m =\gamma = 0.1g$ and $\omega_a=8g$ (solid lines). The oscillator is initialized in a thermal state with an average number of phonons equal to the one of the corresponding thermal bath. From bottom to top (considering the steady-state values): $n_{\sf th} = \{ 0.2, 0.3, 0.4, 1, 3, 5 \}$. Notice that the curves corresponding to the three lower values of $n_{\sf th}$ are almost superimposed. The dashed blue line shows the time evolution of $\Delta\hat{x}_R^2$ for the effective Hamiltonian (its value does not depend on the number of thermal phonons $n_{\sf th}$). 
 {\bf (b)}: Time evolution of the variance $\Delta\hat{x}_1^2$ in dB-scale when
 the feedback protocol is implemented, for the same values of the parameters characterizing 
 the system. From bottom to top: $n_{\sf th} = \{ 0.2, 0.3, 0.4, 1, 3, 5 \}$. }
 \end{figure}

Finally we assess the effects that the bath temperature has on the squeezing performance. As in the zero-temperature case, we observe that the feedback-assisted scheme is able to obtain results qualitatively similar to those achieved through Eq. (2), even for moderate values of the detuning, where the non-assisted protocol fails. In particular, the behaviour of the renormalised variance $\Delta\hat{x}^2_R$ is only slightly affected by the temperature of the bath, which is evidence of the similarity of performance between the feedback-assisted scheme and the ideal one, which is indeed independent of $n_{\sf th}$. Fig. \ref{ThermSqFeed} {\bf (a)} shows {\it de facto} insensitivity to the bath temperature for any value of $n_{\sf th}<0.5$ and only small deviations from the zero-temperature case for larger values of such parameter. 
Squeezing below the vacuum limit, on the other hand, can be achieved only for $n_{\sf th} < 0.3$, as observed in Fig. \ref{ThermSqFeed} {\bf (b)}.

 \par
 
 \section{Analysis of the qubit survival probability}

\begin{figure}[t!]
\centering
\centering{\bf (a)}\hskip6cm{\bf (b)}\\
\resizebox{.35\columnwidth}{!}{\includegraphics{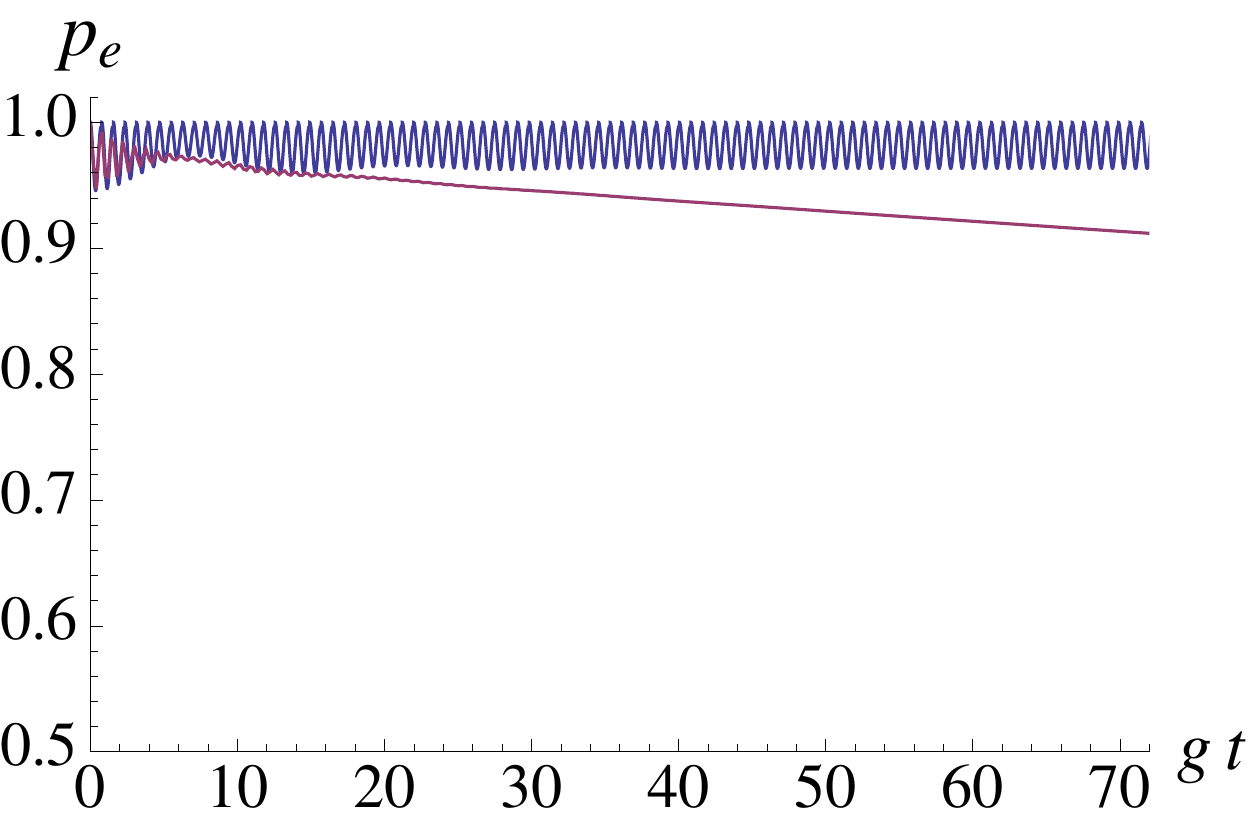}}
\resizebox{.35\columnwidth}{!}{\includegraphics{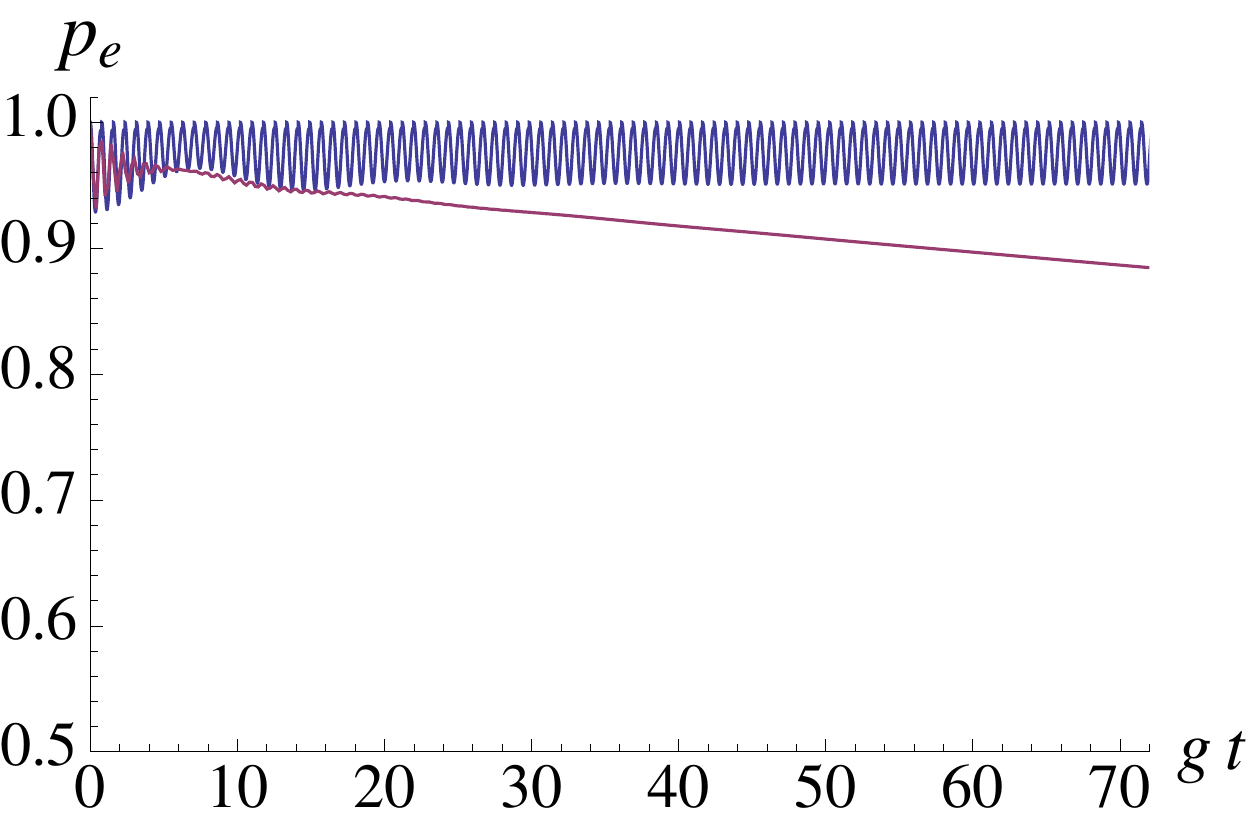}}\\
\vskip0.1cm
\centering{\bf (c)}\hskip6cm{\bf (d)}\\
\resizebox{.35\columnwidth}{!}{\includegraphics{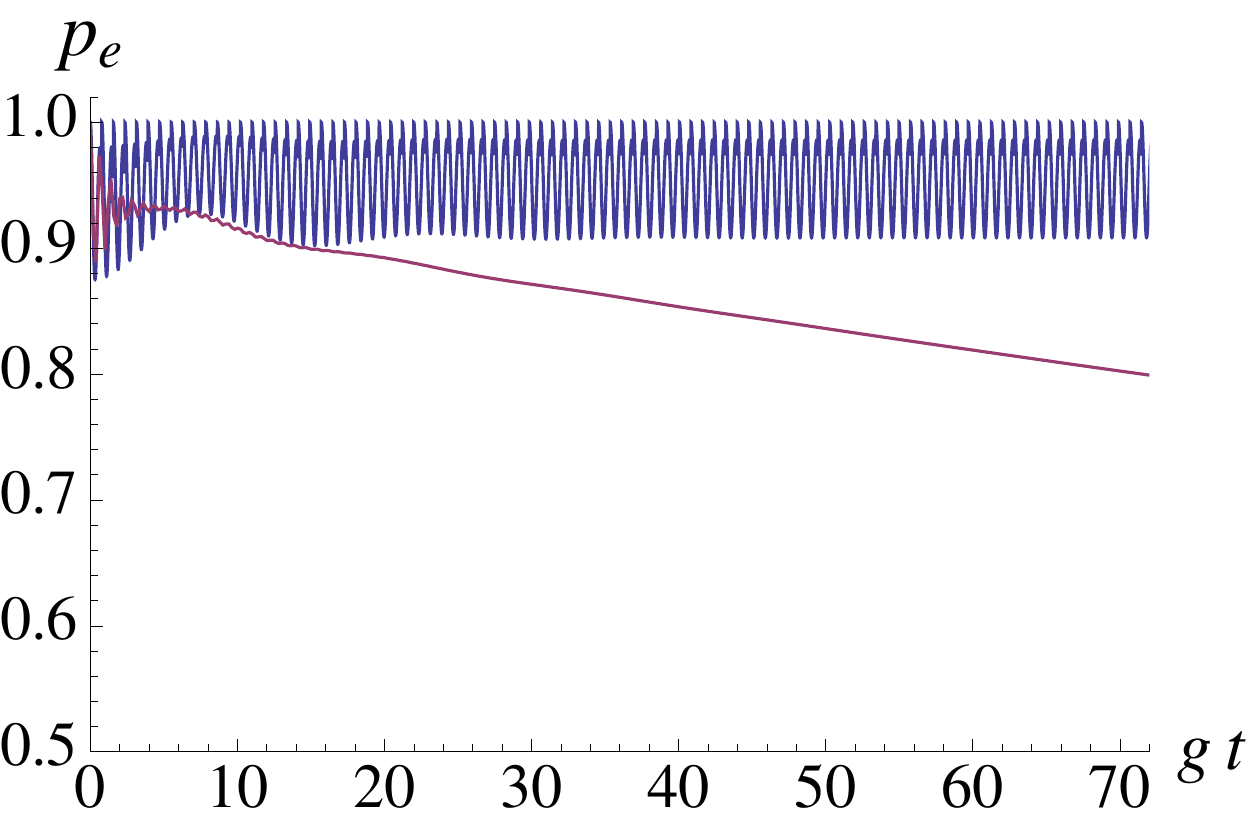}}
\resizebox{.35\columnwidth}{!}{\includegraphics{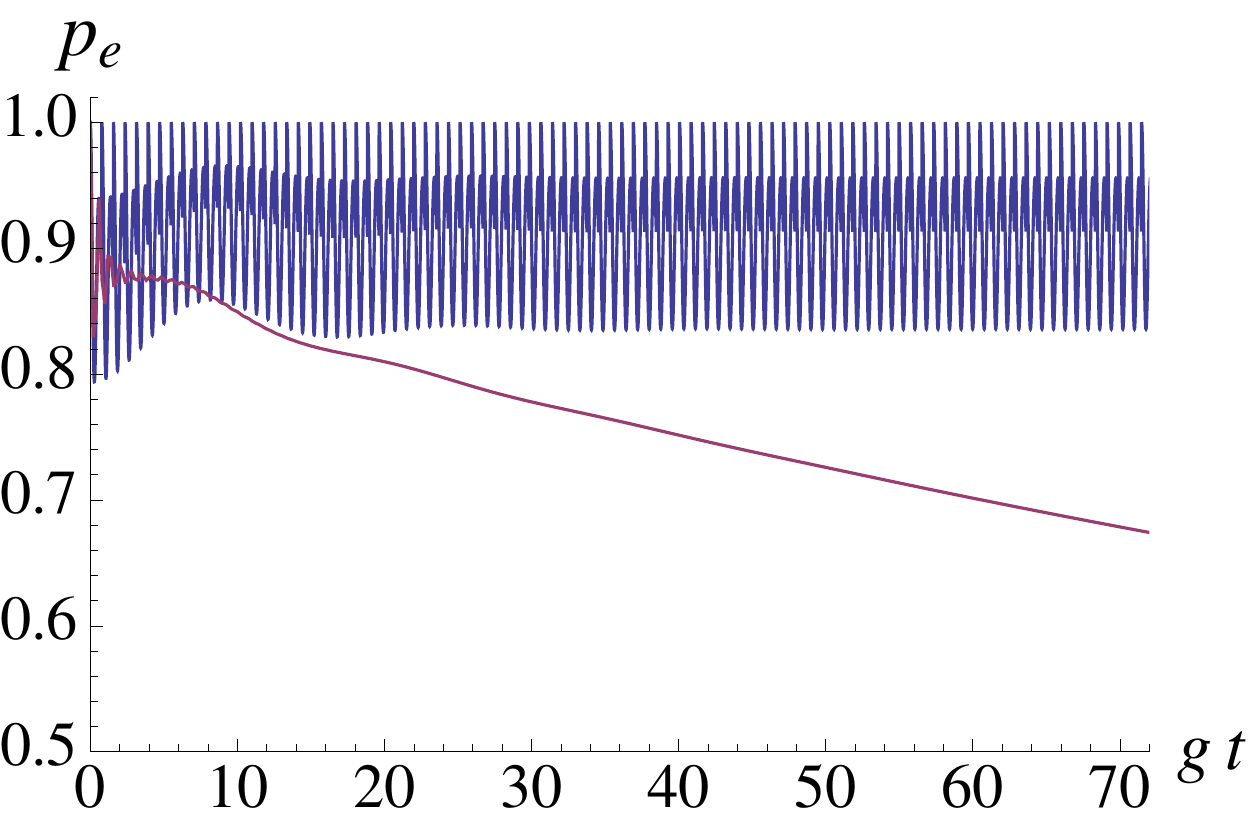}}
\caption{\label{f:pE} Probability $p_e$ against the dimensionless time $g t$ for initial thermal states of the oscillator and corresponding bath with $n_{\sf th}=\{0, 0.2, 1,3\}$ [panel {\bf (a)}, {\bf (b)}, {\bf (c)} and {\bf (d)} respectively]. Other parameters have the same values as in Fig.~\ref{ThermSqFeed}. The numerical simulation with repeated measurements (blue curve) keeps $p_e$ at large values at the steady state, thus enabling the squeezing of the oscillator. Differently, the purple curve (numerical simulation without repeated measurements) shows a decreasing $p_e$. The corresponding steady state exhibits no squeezing.
}
\end{figure}

 As we pointed out above, without the feedback loop no steady-state squeezing can be achieved. Following the discussions made above on the working principles of our protocol, a significant figure of merit for the performance of the squeezing mechanism is embodied by the excited-state survival probability $p_e$ of the qubit. This is plotted in Fig. \ref{f:pE} for four different values of $n_{\sf th}$. The figure reveals that, with our feedback-assisted protocol, the probability of excitation of the qubit is always kept very close to 1, whereas it quickly fades when the protocol is not used. Upon inspection of Eq.~(\ref{H2}), we realise that for a qubit prepared in $\ket{g}$, the harmonic oscillator would be effectively squeezed in a direction opposite to that corresponding to the case of its initialisation in $\ket{e}$. Therefore, if the qubit is not maintained in its excited state, squeezing along opposite directions in phase space is performed, leading to a steady-state with large fluctuations in the quadratures. This ultimately leads to the washing out of the effective mechanism. 
 
More quantitatively, while for small values of $n_{\sf th}$ (i.e. in cases such that quantum squeezing is expected at steady-state), the qubit survival probability is kept by the feedback protocol at values larger than $95\%$, a thermal bath enforces lower values of such probability. As a consequence, no quantum squeezing is obtained. However, the difference with the case where no feedback is implemented is evident, thus leaving room for thermomechanical squeezing.

\section{Conclusions}

We have proposed a feedback-assisted protocol for the steady-state squeezing of a harmonic oscillator. The protocol requires only a limited degree of control over the system , and is thus close to the current experimental state of the art. Contrary to procedures based on the time-controlled interaction between the qubit and the oscillator, our proposal is resource-efficient, as it is based on an always-on interaction that does not need to be tuned. {It is  interesting to compare the  performance of our scheme to the case of parametric driving and driven dissipative architectures. The steady state of parametrically driven oscillators can be squeezed by at most 3 dB before entering self-oscillatory regimes~\cite{Milburn}. When compared to such limit, our scheme is found to perform very well, achieving a steady-state reduction of $\Delta\hat x^2_1\sim 2$ dB. At short evolution times, we can achieve values surpassing this performance and comparing well with schemes based on amplitude-modulation of the optical driving of mechanical devices~\cite{Mari}. The combination of continuous quantum measurements and closed-loop feedback operated on the oscillator~\cite{Ruskov}, or the combination of detuned parametric driving and oscillator position measurements~\cite{Szorkovszky} can surpass the 3 dB steady-state bound (and thus beat our scheme). However, this is achieved at the price of nearly ideal (quantum non-demolition) measurements and challenging feedback mechanisms on the oscillator. Squeezing values well beyond the 3 dB limit can be achieved, dynamically, using multi-tone drivings and clever reservoir engineering~\cite{Kronwald}, or squeezed drivings~\cite{Huang}. Such proposals require the engineering of the environmental system, and it remains to be seen whether replacing this pre-requisite with the use of the feedback mechanism discussed here would actually ease the achievement of mechanical squeezing.  

While this point is best addressed when explicitly designing an experimental setup, and is thus beyond the scopes of the present proposal, we would remark that our scheme can be applied to a range of experimental situations, leaving at the same time room for interesting extensions addressing the area of dissipative quantum state engineering~\cite{dissipative} of harmonic motion, where one could achieve qubit-assisted squeezing in the state of the oscillator.

\section*{Acknowledgments} 

We are grateful to  N. Lambert, M. S. Kim, A. Serafini, and T. Tufarelli for fruitful discussions. This work was supported by the UK EPSRC (through grants EP/I026436/1, EP/G004759/1 and EP/K026267/1), the Alexander von Humboldt Stiftung, the John Templeton Foundation (grant ID 43467) and MIUR (grant FIRB ``LiCHIS'' - RBFR10YQ3H).

\end{document}